\documentclass[11pt,cites,graphicx,hyperref]{article}

\usepackage{graphicx}

\usepackage{hyperref}

\title{Antiferroelectric liquid crystal model with ions diffusion.}

\author{P.L. del Castillo,$^1$ P.L. Lucas,$^2$ N. Bennis,$^3$ A. Spad\l o$^4$ and D. Rodriguez-Perez,$^5$}

\begin{document}

\maketitle

{
$^1$ Department of Mathematics, CEIPS. El Molar n$^\circ$ 2, C/San Isidro 2, El Molar,  E-28710 Madrid, Spain. E-mail: antiferroelectrica@yahoo.es\\
$^2$ Department of Mathematics, IES. Fco. G\'\i ner de los R\'\i os, Carretera Alcobendas-Barajas s/n, Alcobendas, Madrid, Spain. E-mail: selairi@bluebottle.com\\
$^3$ Dpt. Tecnolog\'ia Fot\'onica, ETSI Telecomunicaci\'on, Universidad Polit\'ecnica de Madrid, Ciudad Universitaria, E-28040 Madrid, Spain\\
$^4$ Institute of Chemistry, Military University of Technology, ul. Kaliskiego 2, 00-908 Warsaw, Poland\\
$^5$ Dpto. de F\'\i sica y Matem\'atica de Fluidos, Facultad de Ciencias, UNED, Senda del Rey s/n, 28040 Madrid, Spain\\
}

\abstract{\noindent Antiferroelectric liquid crystals can be considered as a promising alternative to nematic mixtures in the area of microdisplays. Switching behaviour of the molecules has been modelled as two adjacent smectic layers. However, some studies made so far reveal that electrooptical response can be seriously affected by ion content of the test cells, specially in the area of asymmetric cells. Regarding such influence, both molecules and ions have been studied in the present work under the influence of an external electric field. The mechanisms governing ionic behaviour considered so far in the simulations have been diffusion, which was dealt through the Nerst-Planck equation, and electrostatic interaction, which mixes ionic and antiferroelectric liquid crystal effects by means of Gauss equation. In addition, voltage pulses applied to the simulated antiferroelectric liquid crystal cell, have been considered under different waveforms for possible driving schemes in display applications.

Results show time evolution of ions within the cell, under different external applied electric fields and their delocalization from alignment layers when high frequency AC pulses are included in the structure of the switching waveform.}




\section{Introduction.}

\label{intro}

Liquid crystals are becoming crucial components in most flat panel display applications. Nowadays, lots of nematic mixtures are much used to develop a wide range of portable and TV displays with different properties, specially those related to contrast and viewing angle. Last advances in research and industrial manufacturing processes are overcoming main difficulties with respect to resolution, size, and temperature range functioning \cite{BRef.1,BRef.2,BRef.3}. Nevertheless, other drawbacks have been observed from experimental characterization of market displays. This is the case of ghost and still images \cite{BRef.4} that can damage the display by constituting shadows which are permanent, or have a lifetime long enough for being detected by human eye. Such behaviour is responsible for a trailing effect clearly observed in video sequences \cite{BRef.5}, and has been related to the slow response time of nematic liquid crystals.

On the other hand, the group of antiferroelectric liquid crystals (AFLCs) is being studied and characterized due to their hysteresis, experimental fast response time and wide viewing angle. In AFCLs, molecules are arranged in layers, called smectic, and can move describing two helical motions in the direction normal to the smectic layers, with a pitch in the order of $1-2 \mu$m\cite{BRef.6}. Such kind of liquid crystals has been often suggested as a good candidate for solving problems related to slow response of nematic mixtures, like image trailing in the area of microdisplays\cite{BRef.5,BRef.6,BRef.7}. Hysteresis exhibited by these materials raises the possibility of developing passive matrix techniques so that the cost/quality ratio could be decreased even more than that of current TFTs made with nematic mixtures. However, AFLC dynamic properties are experimentally found to be seriously affected by changes in the alignment layer. Since these liquid crystals do not have a nematic phase within their phase sequence, most commercial polyimides are useless for their alignment\cite{BRef.8}. Even those that maintain the main characteristics of AFLC response, result in drastic losses of contrast, commonly related to a leakage transmission in the absence of an applied electric field, called pretransitional effect\cite{BRef.9}.

A summary of the three main possibilities for enhancing contrast of antiferroelectric liquid crystal mixtures can be found in \cite{BRef.10}. First attempts were based on chemical changes that produced a new group of antiferroelectric mixtures, called orthoconic liquid crystals \cite{BRef.11,BRef.12}, and are characterized by their $45^\circ$ tilt angle for the molecules within smectic layers. In surface stabilized structures \cite{BRef.13,BRef.14}, orthoconic materials behave as isotropic layers, giving rise to the best dark state between crossed polarizers \cite{BRef.9,BRef.15,BRef.16}. Apart from changes in the driving scheme \cite{BRef.17}, other possible solution for improving cell contrast can be achieved by producing non-conventional changes in alignment layers. An example of this line of research is the sector of asymmetric cells, where dissimilar alignment layers are spread over both substrates, giving rise to shifted hysteresis cycles \cite{BRef.18,BRef.19}. Nevertheless, such hysteresis shift was found to be time-dependent as described in \cite{BRef.20}. The main explanation for hysteresis instability lays on ion content of the cells, as well as their generation and motion within the alignment layers. Ion content mainly comes from synthesis process of liquid crystal mixtures \cite{BRef.21} or from dissociations impinged by voltage pulses used in driving. Ion effects in nematic liquid crystal mixtures have been extensively studied by ``Liquid Crystals and Photonics'' group from Ghent University \cite{BRef.22,BRef.23}. Their works have shown effects like lateral ion transport parallel to alignment layers \cite{BRef.4,BRef.24}, among others.

The drift-diffusion phenomenon of ions has been also studied in other articles, by means of an equivalent three elements circuit\cite{BRef.46}. Authors relate circuit elements to basic material parameters, specifying the validity range of the circuit in each case, with applications in solid state electrochemistry systems.

A model of ion organization in the steady state under the influence of an external electric field within a nematic liquid crystal cell, can be found in the works of Ionescu, Barbero et al. \cite{BRef.26,BRef.27}. They explain how switching is enhanced for cathodic polarization while inhibited for anodic, in terms of different surface charge densities over LC interfaces for both situations. However, nothing is reported about ion evolution and molecular movement in dynamic behaviour, that is, when a sequence of different voltage pulses at video frequency is applied to the liquid crystal sample under study.

The present work shows the results of a simulation that considers simultaneously ion behaviour and molecular movement under the application of certain combinations of voltage pulses. Ions motion has been described with the Nerst-Planck mechanism \cite{BRef.28}, whereas molecules have been dealt with a model described in \cite{BRef.13,BRef.29} by using the Ginzburg-Landau procedure to obtain the motion equations. The main objective is to elucidate how changes included in the waveform used for driving the liquid crystal cell, can be used for producing ion delocalization from alignment layers.


\section{Theoretical.}

Main ion sources within a liquid crystal cell have been studied, as well as the processes involved in the presence of an applied electric field \cite{BRef.30,BRef.31}. Test cells are supposed to be $1.5 \mu$m thick, and have identical alignment layers of nylon 6 over both glass substrates, in a way similar to experimental studies made so far \cite{BRef.32,BRef.33}. An antiferroelectric mixture is considered within such cell in a surface stabilized chevron structure \cite{BRef.34,BRef.35}. Regarding this arrangement, the displacement of cell ion content has been simulated, taking into account phenomenon of diffusion under the influence of an external electric field. Simultaneously, molecular switching was considered \cite{BRef.13,BRef.29}. This scheme has been applied with different combinations of voltage pulses, in what could be a possible waveform for driving the liquid crystal for display applications \cite{BRef.36,BRef.37,BRef.38}.

In the following sections, the main steps for setting this model and solving the equations involved are described, as well as the results of the simulations performed.


\section{Ionic processes in liquid crystal cells.}

Antiferroelectric liquid crystals contain ions  \cite{BRef.30}. Main ion sources of liquid crystals come from\cite{BRef.31}:
\begin{itemize}
\item Liquid crystal synthesis process.
\item Manufacturing protocol of test cells.
\item Molecular dissociation or ion diffusion within alignment layers.
\end{itemize}

\begin{figure}[ht]
\begin{center}
\includegraphics[width=0.65\textwidth]{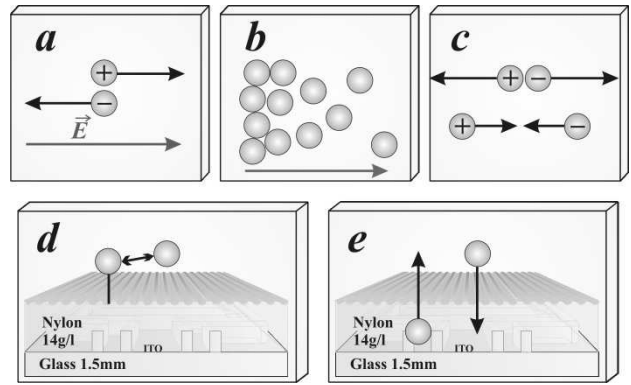}
\caption{Effects on ion content inside a LC cell when an electric field is applied: a) Movement; b)Diffusion; c) Formation or annihilation; d) Adsorption; e) Absorption}
\label{1}
\end{center}
\end{figure}

Typical ion concentration in liquid crystal cells oscillates within the interval $10^{16}-10^{20} \mbox{ions}/\mbox{m}^3$ \cite{BRef.39}. Therefore, a cell $1 \mu$m thick and $1 \mbox{cm}^2$ in area is expected to contain $10^6-10^{10}$ ions. Under the influence of an external electric field applied to the cell, the ion concentration mentioned before evolves being affected by the following combination of effects. Some of them are obviously expected regarding arguments from classical electromagnetic theory, and are schematically represented in the squares of Fig. \ref{1}:

\emph{Motion under the  electric field}: Ion fluxes, $J$, are represented by:
\begin{equation}
J=Cq\mu E
\end{equation}
Where $C$ is the ion concentration, $q$ the ionic charge , $E$ the electric field, and $\mu$ the ion mobility.\\

\emph{Diffusion}: This mechanism and motion mentioned before are linked by the Nernst-Planck equation \cite{BRef.28}:
\begin{equation}
\frac {\partial C} {\partial t}=D\frac {\partial ^2 n} {\partial x^2}-\mu \frac{\partial (nE)} {\partial x}
\end{equation}
Where $D$ and $\mu$ values are assumed to be constant (i.e. the medium is taken as homogeneous).

\emph{Recombination-Generation}: Collisions of ions and molecular dissociations contribute to the temporal evolution of both positive $C_+$ and negative $C_-$ ion concentrations, that can be expressed as follows \cite{BRef.40}:
\begin{equation}
\frac {\partial C_+} {\partial t}=\frac {\partial C_-} {\partial t}=\alpha C_0-\beta C_+C_-
\end{equation}
Where, $C_0$ is molecular concentration, whereas $\alpha$ and $\beta$ are respectively generation and recombination constants.

\emph{Trapping of ions at the alignment layers}: Trapped ions form a layer over alignment surfaces, so that a screen effect over external electric field will be produced. Higher concentrations of free ions ($C$) within the cell are expected to speed up the growth of trapped ions layer. Mathematically, this relation could be expressed as follows:
\begin{equation}
\frac {\partial C_t}{\partial t}=-\frac {\partial C}{\partial t}=KC-\frac {C_t}{t_t}
\end{equation}
$K$ is proportional to the trapping probability; $C_{t}$, $C$ are respectively the concentrations
of trapped ions and those that have been included into the liquid crystal pool. Finally, $t_{t}$  has the meaning of a characteristic time of liberation \cite{BRef.41}.\\

In general, ions generate an electric field against external electric field. Thus, they contribute to decrease the external imposed field value within the liquid crystal cell. A similar effect happens when AFLC polarization is considered. Both effects will be taken into account in the model described in the present work.


\section{One-dimensional ionic model within a LC cell.}
    
A liquid crystal cell behaves as a capacitor, being the antiferroelectric liquid crystal layer the dielectric material. The glass substrates, covered by the semiconductor ITO (Indium-Titanium-Oxide), and nylon 6 as alignment layer, will act as electrodes for the mentioned capacitor. In the next section, a mathematical formalism for ions and molecules will be expounded. Main effects under study are diffusion and drift caused by the external electric field.


\subsection{Ionic diffusion.}
    
Ion diffusion can be described in the framework of the Nernst-Planck equation \cite{BRef.28}:
\begin{equation}
\label{Nernst-Planck}
\frac {\partial C_{\pm}} {\partial t}=-\frac {\partial} {\partial x} \left( -D\frac{\partial C_{\pm}}{\partial x} \mp \mu zeC_\pm\frac {\partial V}{\partial x}\right)
\end{equation}
Being $C_{+}$, $C_{-}$ the concentrations of positive and negative ions, $\mu$ the ion mobility within the liquid crystal, $C_0$ the total ionic concentration, $D$ the ionic diffusivity , $ze$ the ionic charge.\\
In addition, the following quantities will be defined:\\
\begin{equation}
\begin{array}{ccc}
\displaystyle{c=\frac{C_++C_-}{2C_0} }&\displaystyle{\rho=\frac{C_+-C_-}{2C_0}}&\displaystyle{\mu=\frac D {k_BT}}
\end{array}
\end{equation}
$\rho$ is related to the charge concentration.\\
$C$ is the total (positive and negative) ionic concentration.\\
$\mu$ is the ionic mobility expressed, by means of Einstein's relation, in terms of the temperature $T$ of the cell.\\
The Nernst-Planck equations can be rewritten in terms of the new quantities $\rho$ and $C$ as follows:
\begin{equation}
\left.
\begin{array}{l}
\displaystyle{ \frac {\partial c}{\partial t}=-\frac{\partial}{\partial x}\left(-D\frac{\partial c}{\partial x} -\frac {Dze}{k_BT}\rho \frac{\partial V}{\partial x} \right) }\\
\displaystyle{ \frac {\partial \rho}{\partial t}=-\frac{\partial}{\partial x}\left(-D\frac{\partial \rho}{\partial x} -\frac {Dze}{k_BT}c \frac{\partial V}{\partial x} \right) }\\
\end{array}
\right\}
\label{eq7}
\end{equation}


\subsection{Molecular movement.}
In order to describe molecular movement, the model developed by Sabater et al. \cite{BRef.13,BRef.29} will be used. This model considers the following most relevant interactions between molecules:\\
Antiferroelectric interaction:\\
\begin{equation}
\displaystyle{f_c=K_p\vec{P_1}\cdot \vec{P_2}=K_p P_S^2 \cos(\varphi_2-\varphi_1)} 
\end{equation}
Polarization:\\
\begin{equation}
\displaystyle{ f_{pol}=P_S\cos{\delta}\left( \sum_{i=1}^2 \cos \varphi_i \right)\frac {\partial V}{\partial x} }
\end{equation}
Dielectric energy:\\
\begin{equation}
\begin{array}{rl}
\displaystyle{f_{Diel}=}&\displaystyle{-\frac 1 2 \sum_{i=1}^2 \left[ \epsilon_1+\delta\epsilon\cos^2\delta\cos^2\varphi_i+\right.}\\
 \\ & \displaystyle{\left.\Delta\epsilon(\cos\theta\sin\delta-\sin\theta\cos\varphi_i\cos\delta)^2  \right]\left(\frac {\partial V} {\partial x}\right)^2}
  \end{array}
\end{equation}
Elastic energy:\\
\begin{equation}
\begin{array}{l}
 \displaystyle{f_E=
 \left(
 	\frac {A'} {2} \cos^2 \delta + \frac {B'} {2} \sum_{i=1}^2 \sin^2 \varphi_i -\frac {C'} {2} \sin 2\delta \sum_{i=1}^2 \sin \varphi_i
 \right)\left(\frac {\partial \delta} {\partial x}\right)^2}\\
 \displaystyle{
 +\frac {B'} {2} \sum_{i=1}^2 \left(\frac {\partial \varphi_i} {\partial x}\right)^2 + C'\cos^2\delta
\left( \sum_{i=1}^2 \cos \varphi_i \frac {\partial \varphi_i} {\partial x} \right)\frac {\partial \delta} {\partial x}
}\\
\displaystyle{
+\frac {D'} {2}\cos\delta \sum_{i=1}^2 \sin 2\varphi_i\frac {\partial \delta} {\partial x} + D'\sin\delta \sum_{i=1}^2 \frac{ \partial \varphi_i} {\partial x}
}\\
\displaystyle{
+\frac {L'} {2}\left(  \frac 1 {\cos^2\delta} -  \frac 1 {\cos^2\delta_{chevron}}   \right)^2}
\end{array}
\end{equation}
Total\ energy:\\
\begin{equation}\label{energia_total}
f=f_c+f_{pol}+f_{Diel}+f_E
\label{eq12}
\end{equation}

\begin{figure}[ht]
\begin{center}
\includegraphics[width=0.5\textwidth]{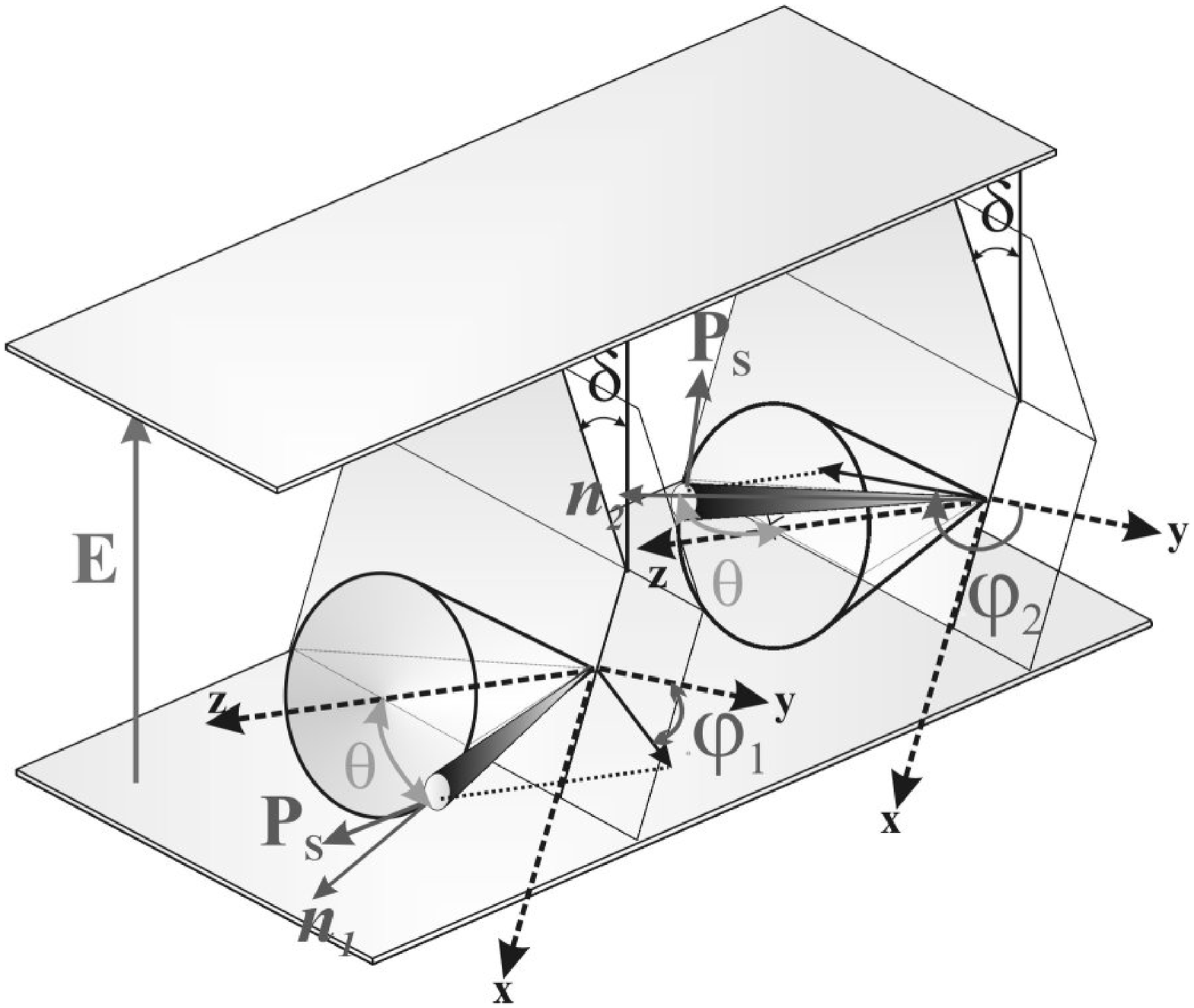}
\caption{Parameters of molecular model in 1D.}
\label{2}
\end{center}
\end{figure}

In previous expressions, $\delta$ is the \textit{chevron} angle, $\delta_{chevron}$ is the chevron angle in absence of external electric fields, $\varphi_{i}$ are the azimuthal angles for molecular positions within smectic layers, $P_S$ the material polarization, $A'$, $B'$, $C'$, $D'$, $L'$  the liquid crystal elastic constants, and $\epsilon_{i}$ its dielectric constants. These parameters are shown in Fig. \ref{2} for the one-dimensional case. On the other hand, $\Delta\epsilon$ and $\delta\epsilon$, birefringence and biaxiality, respectively, are used to define a new parameter $v$ for simplicity within the equations:
\begin{equation}
v=\Delta\epsilon\sin^2\theta-\delta\epsilon
\end{equation}

Interaction energies (\ref{energia_total}) can be linked by means of \emph{Ginzburg-Landau} \cite{BRef.42}:
\begin{equation}
\lambda\frac{\partial \varphi_i} {\partial t}=-\frac{\partial f}{\partial \varphi_i}+\frac{d}{d x}\frac {\partial f} {\displaystyle{ \left( \frac {\partial \varphi_i} {\partial x} \right)} } 
\label{eq14}
\end{equation}


\subsection{Molecular and ionic interaction: Gauss equation.}
Ions and molecules create internal electric fields by the mechanisms mentioned so far. Interactions from both sides are considered simultaneously in the equation of Gauss:
\begin{equation}
\vec{\nabla}\cdot \vec{D}=\vec{\nabla}\left(\epsilon_0\vec{E}+\vec{P}\right)=\rho2C_0ze
\label{eq15}
\end{equation}
Where $\vec{P}$ as polarization vector is made of $\vec{P_P}$, the spontaneous polarization, and $\vec{P_{X}}$, a term related to the dielectric susceptibility as follows:
\begin{equation}
\begin{array}{l}
\displaystyle{\vec{P_P}=P_S\sum_{i=1}^2\left( -\cos\varphi_i\cos\delta,-\sin\varphi_i, \cos\varphi_i\sin\delta\right) }\\
\vec{P_X}=X_{xx}\epsilon_0\vec{E}
\end{array}
\end{equation}


\subsection{Boundary conditions.}
In the process of solving Gauss equation for elucidating ions and molecules behaviour in the presence of an external electric field, three effects should be taken into account, as boundary conditions:\\
\emph{Surface interaction}, which is considered for molecules in the proximity of electrode layers:
\begin{equation}
\begin{array}{c}
\displaystyle{
	\frac {\partial f_0} {  \displaystyle{ \left( \frac {\partial \varphi_i} {\partial x} \right) } } \pm \frac {\partial f_{surf}} {\partial \varphi_i}=0
}\\
\displaystyle{
	f_{surf,up,i} = -g_{up}\left[c_{up}\exp\left( -\alpha\sin^2\frac {\varphi_i-\varphi_{up}} {2}\right) + (1-c_{up})exp\left( -\alpha\cos^2\frac {\varphi_i+\varphi_{up}} {2}\right)\right]
}\\
\displaystyle{f_{surf,down,i} = -g_{down}\left[c_{down}\exp\left( -\alpha\sin^2\frac {\varphi_i-\varphi_{down}} {2}\right)  + (1-c_{down})exp\left( -\alpha\cos^2\frac {\varphi_i+\varphi_{down}} {2}\right)\right]}\\
\end{array}
\label{eq17}
\end{equation}
Being $g_i$, $\beta_i$ constants referred to the alignment layers over both \emph{up} and \emph{down} glass substrates of the test cell \cite{BRef.43}.\\
\emph{Ion confinement within the cell}\cite{BRef.28}:
\begin{equation}
\begin{array}{c}
\displaystyle{ \frac {\partial c} {\partial x} + \frac {ze} {k_B T}\rho\frac{\partial V}{\partial x}=0 }\\
\displaystyle{ \frac {\partial \rho} {\partial x} + \frac {ze} {k_B T}c\frac{\partial V}{\partial x}=0 }\\
\end{array}
\label{eq18}
\end{equation}
\emph{Voltage value in the walls}, that should coincide with values externally imposed by the driving scheme:
\begin{equation}
V=V_\pm \mp \lambda_S \frac {\partial V}{\partial x}
\label{eq19}
\end{equation}


\subsection{Simulation parameters.}
Simulations have been done for the AFLC CS-4001 \cite{BRef.44}. Parameter values for this liquid crystal mixture are shown in Table \ref{tabla_parametros_simulacion}, as well as approximations for constants related to surface boundary conditions. Some of them were used by Sabater \cite{BRef.13,BRef.29,BRef.43} in his studies for the AFLC CS-4000.

\begin{table}
\begin{center}
\caption{\label{tabla_parametros_simulacion} Simulation parameters.}
\begin{tabular}{|c|c|c|} 
\hline
Diffusion & CS-4001 material constants & Material constants\\
\hline
$T=308K$ 				& $A'=10^{-11}N$ 		& $K_p=10^8$ \\
$D=10^{-11}m^2/s$ 			& $B'=10^{-12}N$ 		& $\lambda=0.26$ \\
$q=1.6\cdot 10^{-19} C$ 		& $C'=10^{-12}N$		& $\alpha=5$ \\
$\epsilon(\textrm{Stern layer})=0.01$ 	& $D'=10^{-12}N$ 		& $g_{up}=5\cdot 10^{-5}N/m$ \\
$C_0=10^{16}\ ions/m^3$ 		& $L'=25000 J/m^3$		& $g_{down}=5\cdot 10^{-5}N/m$ \\
 					& $\theta_{semicone}=0.472984 rad$& $\varphi_{up}=2.8274 rad$ \\
 					& $\delta_{chevron}=0.418879 rad$ 	& $\varphi_{down}=0.314159 rad$ \\
 					& $\epsilon_x=4.425\cdot 10^{-11} C^2/Nm^2$ & $c_{up}=0.5$ \\
 					& $\Delta\epsilon=-4.425\cdot 10^{-12} C^2/Nm^2$ & $c_{down}=0.5$\\
 					& $\delta\epsilon=-8.51\cdot 10^{-13}  C^2/Nm^2$ & \\
					& $P_S=79.7\cdot 10^{-5} C/m^2$ & \\
\hline
\end{tabular}
\end{center}
\end{table}


\section{Results.}
\begin{figure}[!h]
\begin{center}
\includegraphics[width=0.49\textwidth]{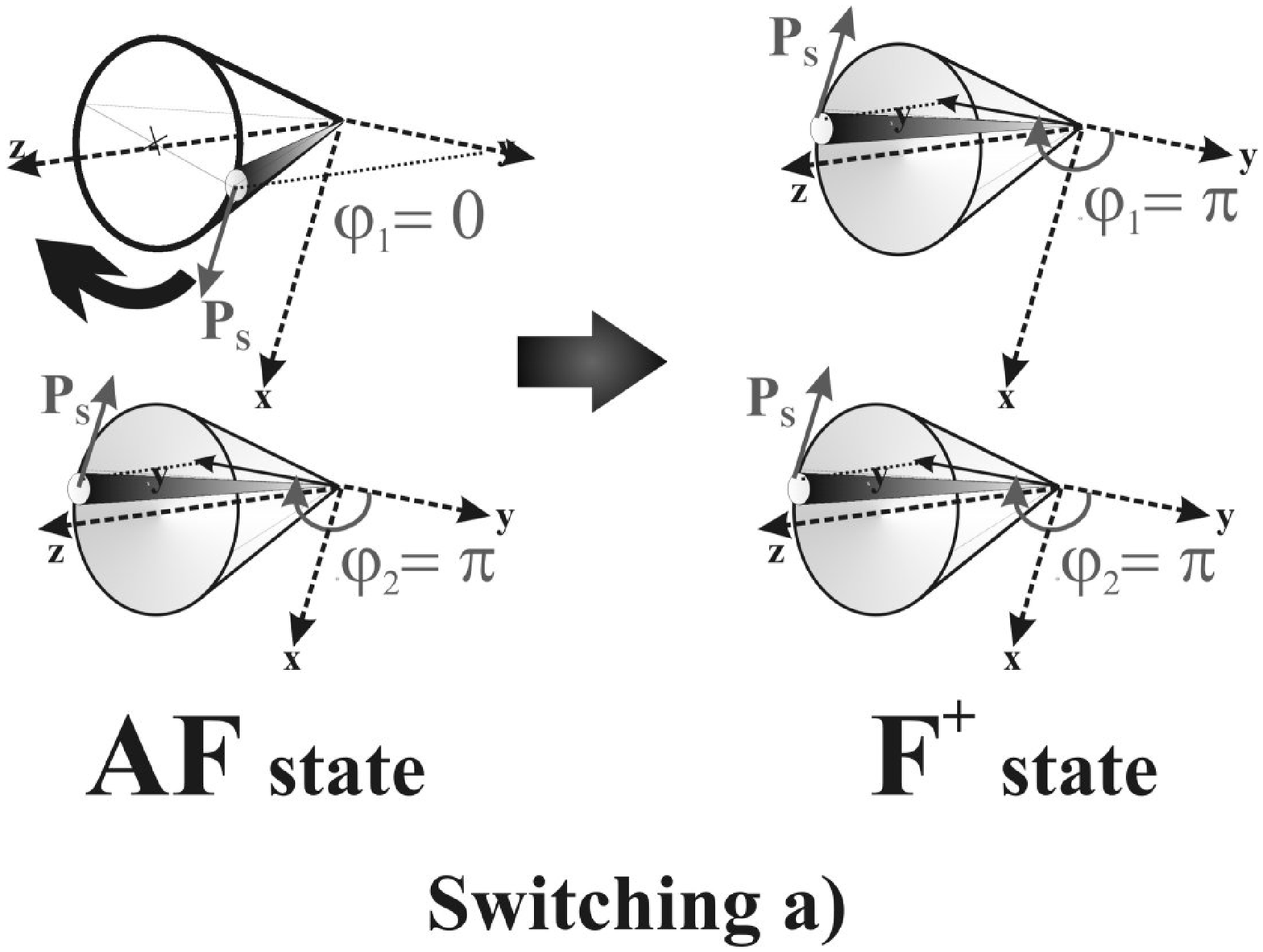}
\includegraphics[width=0.49\textwidth]{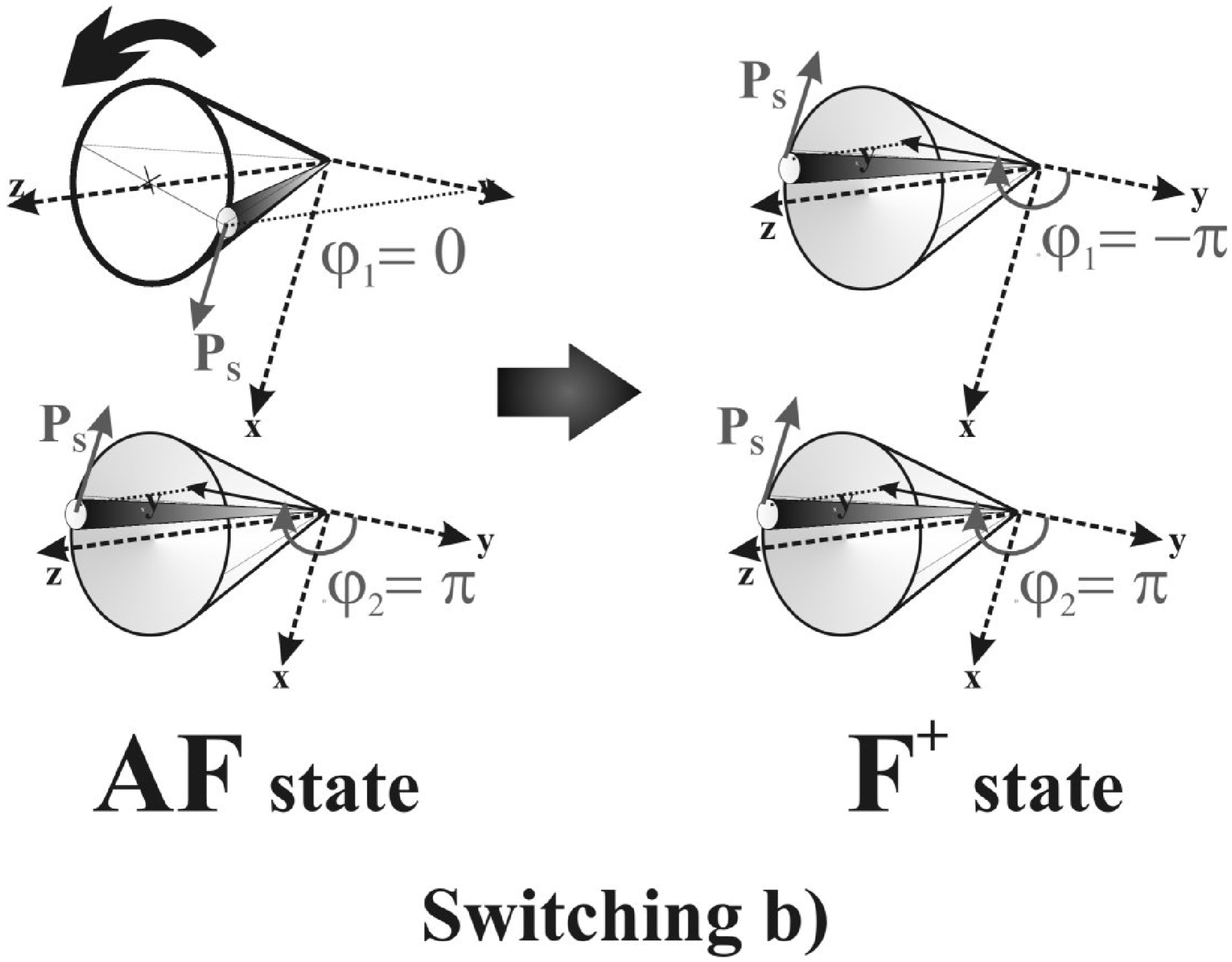}
\caption{Switching process from the antiferroelectric to one of the ferroelectric states. Cases a) and b) are the same, although mathematically they can be considered as different situations.}
\label{5}
\end{center}
\end{figure}

The model formed by the equations \ref{eq7},\ref{eq12},\ref{eq14} and \ref{eq15} and boundary conditions \ref {eq17}-\ref{eq19} was implemented using a finite differences scheme \cite{BRef.45}. Figs. \ref{3}--\ref{6} show a summary of the results obtained for the simulations run with different combinations of voltage pulses. In all cases ion concentration was fixed to $10^{16} \mbox{ions}/\mbox{m}^3$ . Information obtained has been classified in four groups of results, which include molecules and charge evolution data. Regarding external voltage, two waveforms were tested: a first bipolar kind and a simple step with a strategy for speeding up the relaxation from a switching state to the antiferroelectric arrangement.
Finally, a high frequency AC pulses combination was tried to study the effect of fast polarization changes induced by the external applied voltage in the liquic crystal response. In all cases, width and amplitude of pulses used were chosen within the intervals of typical switching conditions experimentally observed for the CS-4001 AFLC mixture in surface stabilized structures \cite{BRef.37}. In addition, cited waveforms have also shown their utility as possible driving schemes for AFLC in microdisplay applications \cite{BRef.38}.


\subsection{Simple bipolar  waveform.}

\begin{figure}[!h]
\begin{center}
\includegraphics[width=0.5\textwidth]{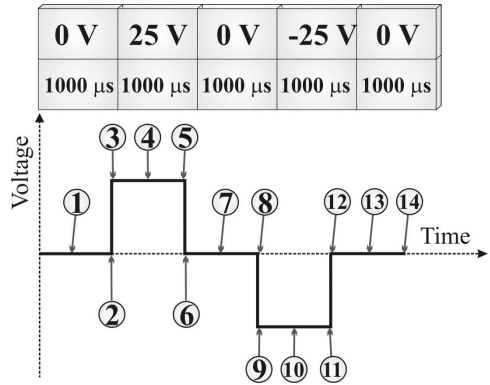}
\caption{Simple bipolar waveform with reset pulse between selection pulse [3-5] and its DC compensation pulse [9-11].}
\label{3}
\end{center}
\end{figure}

\begin{figure}[!h]
\begin{center}
\includegraphics[width=0.72\textwidth]{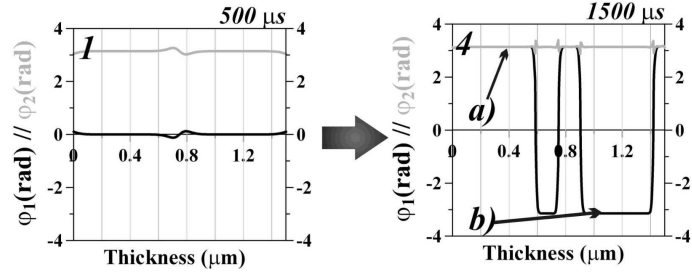}
\caption{Plots of the molecular azimuthal angles in the non-switched (step 1, left), and the switched (step 4, right) configurations. In the last case, the ferroelectric state was reached in two ways (a and b).}
\label{4}
\end{center}
\end{figure}

The waveform is formed by a pre-switching reset pulse ($0$ V) for testing the antiferroelectric arrangement of molecules (step 1 in Fig. \ref{3}). The next pulse is a  $25$ V selection pulse during $1$ ms (steps 2-6 in Fig. \ref{3}). Another reset follows the sequence (steps 6-8), and a DC compensation for the two last pulses finishes the waveform under study (steps 8-14). The data computed from the simulations are grouped as follows:
\begin{itemize}
\item A first group with information about the positions of the molecules. It is given through the azimuthal angles, $\varphi_1$ and $\varphi_2$, which will determine whether the liquid crystal is in one of the ferroelectric states (F+, F-) or in the antiferroelectric one (AF).
\item A second group with information about \emph{chevron} and electric field evolution within the cell.
\item A third group with information about the charge concentration, $\rho$ and how it is influenced by changes in the waveform used for driving. 
\end{itemize}
Thus, it is expected to find AF in the previous reset pulse (step 1 in Fig. \ref{3}), as well as F+ or F- under the selection pulse influence (steps 3-5 in Fig. \ref{3}), that is, the simulation should reproduce such results to show an agreement with experimental results reported by other authors \cite{BRef.37,BRef.38,BRef.44}. Molecular positions can be observed in Fig. \ref{4} for steps 1 and 4 of the waveform proposed in Fig. \ref{3}. An analysis of the graphs shows that the switching state can be achieved in two ways, labeled as a) and b) in the figure. Differences are based on the path followed by the molecules to switch from the AF to the F+ state. All the simulations carried out show that a) or b) election depends on the boundary conditions imposed, as well as on the \emph{chevron} simulated in the middle of the cell.

\begin{figure}[!h]
\begin{center}
\includegraphics[width=0.32\textwidth]{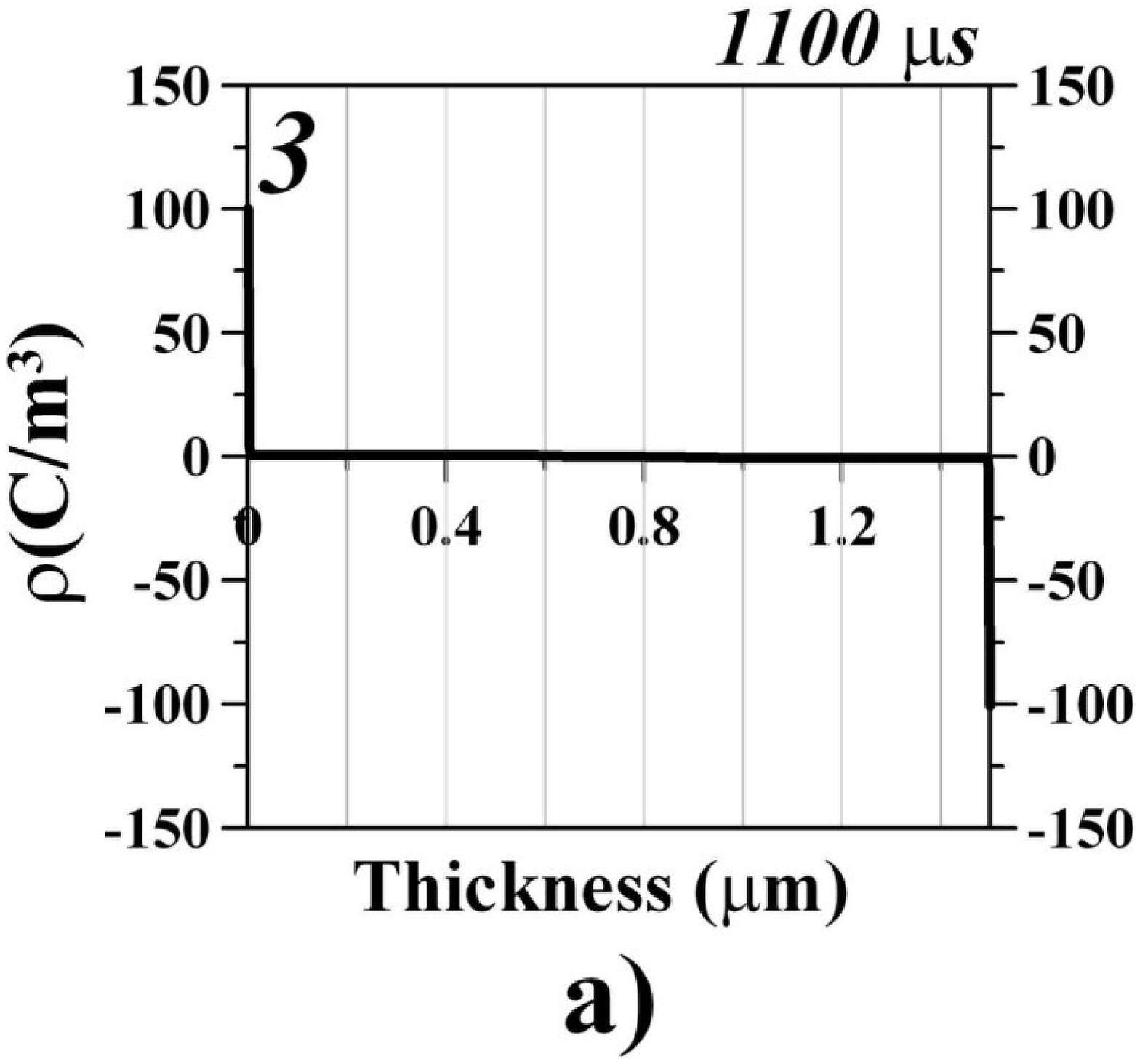}\includegraphics[width=0.32\textwidth]{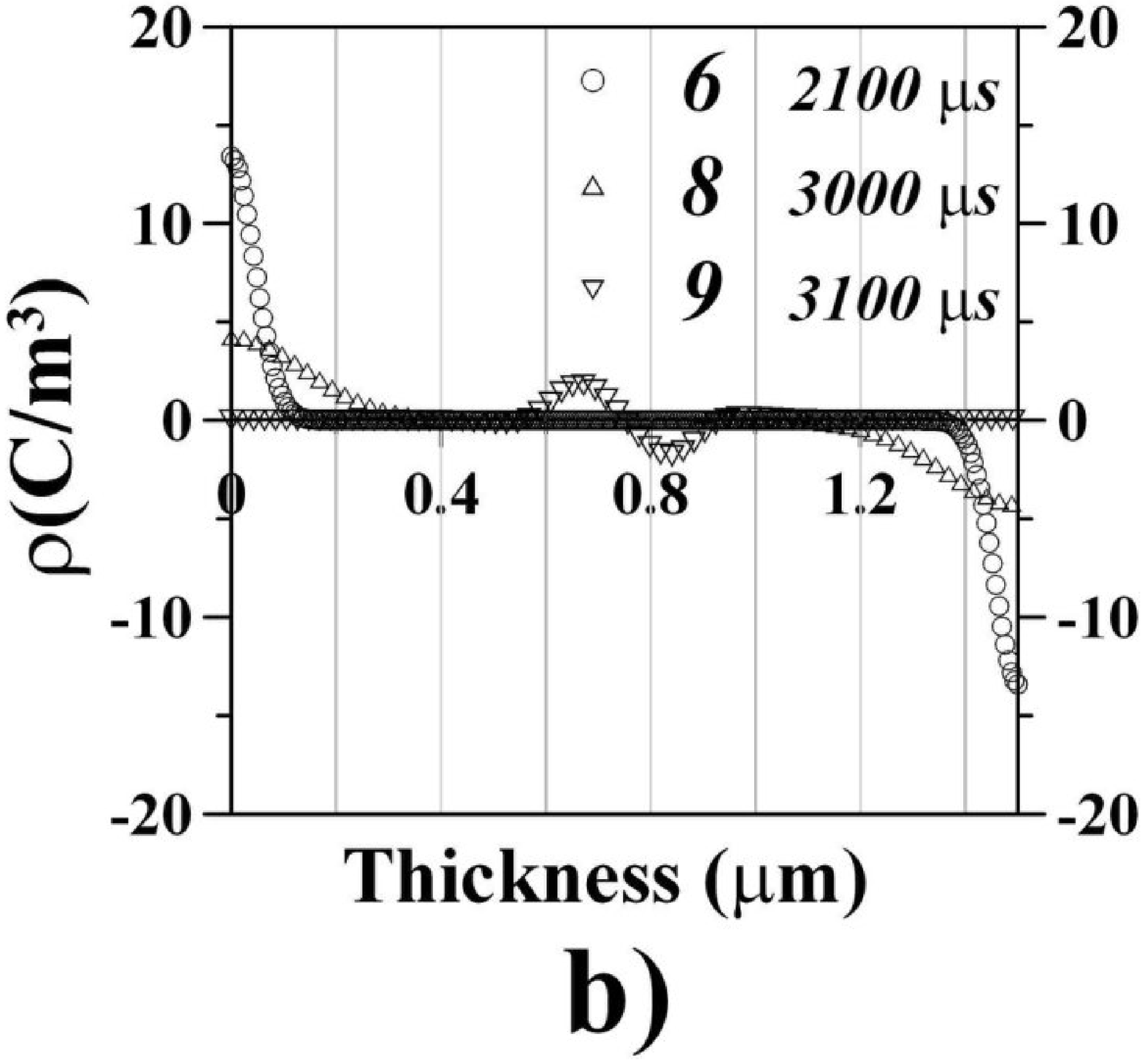}\includegraphics[width=0.32\textwidth]{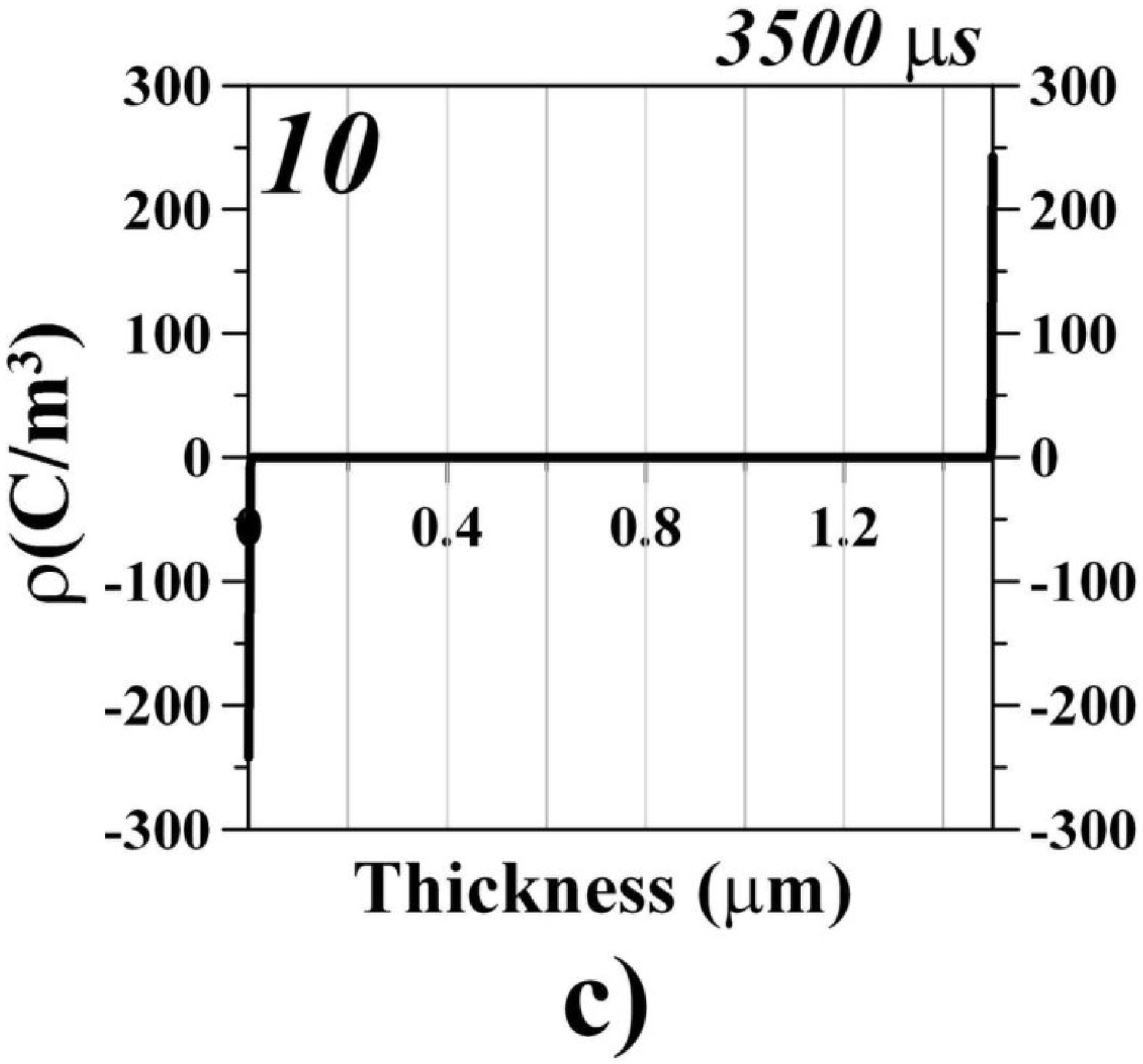}
\caption{Charge density within the cell in different steps of the bipolar waveform applied. Information of the time period is also shown in each graphic: a) Step 3 (1100 $\mu s$); b) Step 6 (2100 $\mu s$), step 8, (3000 $\mu s$), and step 9 (3100 $\mu s$). This plot shows progressive charge delocalization from alignment surfaces. c) Step 10 (3500 $\mu s$).}
\label{6}
\end{center}
\end{figure}

For electric charge density, ion migration towards the walls of the cell is produced under the switching pulse influence. This charge separation is partially maintained during the reset, and can be reversed with a switching pulse of opposite polarity. These results are summarized in Fig. \ref{6}.



\subsection{Biasringlet bipolar waveform.}

\begin{figure}[ht]
\begin{center}
\includegraphics[width=0.35\textwidth]{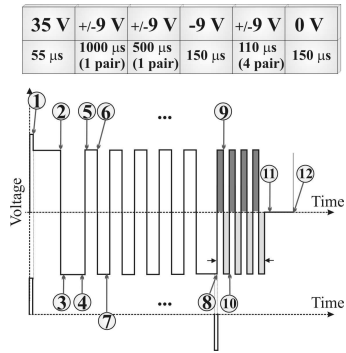}
\caption{Biasringlet bipolar waveform, characterized by secuences of pulses of different polarity in both bias and relaxation sections.}
\label{7}
\end{center}
\end{figure}

\begin{figure}[h]
\begin{center}
\includegraphics[width=0.45\textwidth]{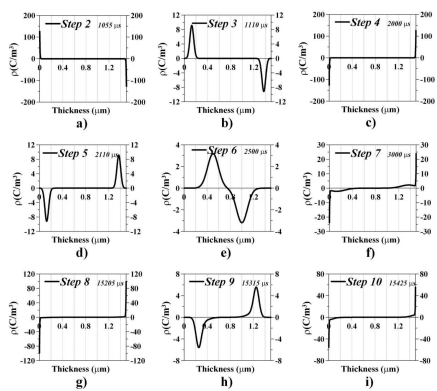}
\caption{Evolution of the net charge density within the cell in strategy steps of the Biasringlet waveform. The step is marked in the upper right corner of each plot, followed by the time elapsed so far in microseconds.}
\label{8}
\end{center}
\end{figure}

The waveform is formed by a selection pulse of $35$ V during $55 \mu$s (step 1 in Fig. \ref{7}), followed by a sequence of high frequency AC pulses. Electrooptical studies made so far \cite{BRef.38} show that the period of the AC scheme in the bias region (for maintaing the transmission, steps 2-16 in Fig. \ref{7}) should be of lower frequency than the one in the section for speeding up the relaxation of the AFLC (steps 16-19). A final reset at $0$ V (step 20) completes the blanking sequence for leading the liquid crystal to the antiferroelectric state. Regarding ion behaviour, fast changes of polarity are expected to produce an effective delocalization from alignment layers. Fig. \ref{8} shows an evolution of the net charge concentration obtained from our simulations, in the main steps of the waveform depicted in Fig. \ref{7}. It has been theoretically proved that fast changes of polarity contribute to charge attenuation and ion delocalization from alignment layers. These results were only found when molecular movement was considered, so can be regarded as a result of the interaction between ions and molecules in switching conditions. An analysis of the plots for azimuts  $\varphi_1$ and $\varphi_2$ reveals that the cell becomes partitioned in multiple regions with different switching behaviour, as is shown in Fig. \ref{5}. This kind of response to external pulses can create a variety of grain boundaries which contributes to ion dispersion specially observed in the plots b), e) and h) of Fig. \ref{8}. In fact, the mentioned delocalization seems to be promoted by an increase in the frequency used for changing the polarity of external voltage pulses, accounting for an attenuation of around a $1$\% of the total charge involved. Molecular relaxation is also faster than that achieved with a simple bipolar waveform, as has been reported by electrooptical studies \cite{BRef.38}.

\section{Conclusions.}

A model of a surface stabilized antiferroelectric liquid crystal cell has been developed that takes into account molecular and ionic motions in the presence of an external electric field. Simulations were performed for different combinations of voltage pulses, in what have been considered valid waveforms for driving these cells in electrooptical studies made so far \cite{BRef.17,BRef.18,BRef.19,BRef.20}. The results first reproduce switching conditions when external voltage achieves threshold and saturation levels. Ion motion creates an internal field opposed to the external one. However, fast changes of polarity in the external field produce ionic delocalization from alignment layers as well as weak net charge attenuation for high frequency AC pulses. Fast changes of external electric field can produce unstable molecular electric fields that interact with ions. These local fields produce ionic concentration bursts ``far away'' from the alignment layers. As this effect does not appear in simulations developed in the absence of molecular movement, ionic delocalization has been related to the dynamical interplay between ions and molecules under high frequency AC pulses. As a result, driving schemes based on fast changes of polarity can be useful in the task of controlling side effects of ion concentration within AFLC cells, as possible delays of the LC response against external voltage pulses, or cell aging caused by driving.

Further studies with other waveforms and different LC constants are under study.

\section{Acknowledgements.}

Thank you very much to Professor Kristian Kneys. His wide knowledge of ions within liquid crystals has motivated the authors of this work to carry out these and other simulations.
Thank you to Professor Ot\'on, and all the members of the CLIQ group.



\end{document}